\documentclass[%
 reprint,
showpacs,preprintnumbers,
 amsmath,amssymb,
aps,
floatfix
]{revtex4-1}

\usepackage{graphicx}
\usepackage{xcolor}
\usepackage{dcolumn}
\usepackage{bm}
\usepackage{hyperref}
\hypersetup{
    colorlinks=true,
    linkcolor=blue,
    citecolor=blue,     
    urlcolor=blue,
}
\usepackage{amsmath}
\usepackage{makecell}

\allowdisplaybreaks

\begin{document}

\title{Isothermal current-driven insulator-to-metal transition in VO$_\text{2}$ through strong correlation effect}

\author{Yin Shi}
 \email{yxs187@psu.edu}
\author{Long-Qing Chen}%
 \email{lqc3@psu.edu}
\affiliation{%
Department of Materials Sciences and Engineering, Pennsylvania State University, University Park, PA 16802, USA
}%

%
%

\date{\today}

\begin{abstract}
Electric current has been experimentally demonstrated to be able to drive the insulator-to-metal transition (IMT) in VO$_2$. 
The main mechanisms involved are believed to be the Joule heating effect and the strong electron-correlation effect. 
These effects are often entangled with each other in experiments, which complicates the understanding of the essential nature of the observations. 
We formulate a phase-field model to investigate theoretically in mesoscale the pure correlation effect brought by the current on the IMT in VO$_2$, i.e., the isothermal process under the current. 
We find that a current with a large density ($\sim 10^1$ nA/nm$^2$) induces a few-nanosecond ultrafast switch in VO$_2$, in agreement with the experiment. 
The temperature-current phase diagram is further calculated, which reveals that the current may induce the M2 phase at low temperatures. 
The current is also shown capable of driving domain walls to move. 
Our work may assist related experiments and provide guidance to the engineering of VO$_2$-based electric switching devices.


\end{abstract}

\maketitle


\section{Introduction}
Insulator-to-metal transition (IMT) in the strongly correlated electron system, vanadium dioxide (VO$_2$)~\cite{Morin59Oxides}, has been attracting widespread attention; it not only provides a platform for fundamental scientific research of strong correlation physics~\cite{Biermann05Dynamical,Zheng15Computation,Brito16Metal}, but also gives rise to novel device applications such as sensitive sensors, Mott field-effect transistors and memristors~\cite{Liao17Ultrafast,Nakano12Collective,Son11Excellent,Driscoll09Phase,Yang11Oxide}. 
Above the transition temperature $T_\text{c} = 338$~K~\cite{Park13Measurement}, VO$_2$ is a rutile (R) metal, while below $T_\text{c}$, it turns into a monoclinic (M1) insulator, at which the resistivity, infrared transmission and eigenstrain change dramatically~\cite{Zylbersztejn75Metal,Chain91Optical}. 
Doping~\cite{Marezio72Structural} or the application of uniaxial stress~\cite{Pouget75Electron} can stabilize another monoclinic (M2) insulating phase.
The IMT can be induced by various external stimuli such as temperature, stress (strain), doping and light~\cite{Morin59Oxides,Pouget75Electron,Marezio72Structural,Cavalleri01Femtosecond}. 
It has been experimentally demonstrated that the IMT can also be triggered by the electric voltage, which is of particular interest owing to its potential application in information technology~\cite{Nakano12Collective,Son11Excellent,Driscoll09Phase,Yang11Oxide,Wu11Electric}.

Although the electric field alone (in an open circuit) can drive the IMT~\cite{Nakano12Collective,Wu11Electric}, the electric current commonly accompanying the electric field (in a closed circuit) may play a role in the electrically triggered IMT~\cite{Leroy12High,Seo11Voltage,Zhou13Voltage,Chae05Abrupt,Taketa77Switching}. 
Unlike in the field-driven IMT that the initial insulating state changes to the equilibrium metallic ground state, in the current-driven IMT the insulating state changes to the nonequilibrium metallic steady state. 
Two dynamic processes will occur: the current will heat up the system through Joule heating effect, and also inject free carriers into the system which will screen the electron-electron repulsion and thus reduce the electron correlation~\cite{Kim04Mechanism,Stefanovich00Electrical,Wegkamp14Instantaneous}. 
The former Joule heating effect can lead to temperature rising above $T_\text{c}$ and thus trigger the IMT simply thermally. 
On the other hand, the latter electron correlation effect brought by the current may delocalize the electrons in the insulating state, thereby induce the IMT as well. 
These two mechanisms are often entangled with each other, which complicates the understanding of the essential nature of the observations on the current-driven IMT. 

Some experiments and simulations employing dc bias and low-frequency voltage pulse supported the Joule heating as the main mechanism for the current-driven IMT~\cite{Simon13Role,Radu15Switching,Lee13Origin,Zimmers13Role,Li16Joule}. 
In particular, using the fluorescence spectra of rare-earth doped micron-sized particles as local temperature sensors, Zimmers \textit{et al.} found that the local temperature of the VO$_2$ sample reaches the transition temperature $T_\text{c}$ as the IMT is induced by a dc current~\cite{Zimmers13Role}. 
Nevertheless, other experiments showed that the transition voltage weakly depends on the thermal dissipation rate and the initial temperature of the VO$_2$ sample, indicating that the IMT is unlikely to be induced by the Joule heating effect~\cite{Joushaghani14Voltage,Yang11Studies}. 
Furthermore, it has been found that the application of a voltage pulse of few volts (accompanied by a corresponding current pulse) switches VO$_2$ from insulator to metal in few or tens of nanoseconds~\cite{Leroy12High,Zhou13Voltage,Chae05Abrupt}. 
This ultrafast switching can hardly be attributed to the Joule heating mechanism, since the time scale of the Joule-heating-induced switching is expected to be at least one order larger than the time scale of the switching observed in the experiments~\cite{Leroy12High,Zhou13Voltage,Chae05Abrupt}. 
Hence, the ultrafast switching must be driven primarily by the electron correlation effect coming along with the current.

Despite of these experimental observations of the current-induced ultrafast switching, the theoretical modeling of this phenomenon is still lacking, which is however desired for understanding the phenomenon and providing guidance to experiments and device applications. 
Previously we have formulated a phase-field model to describe the IMT in VO$_2$ with the thermodynamics described by a Landau potential as a function of structural order parameters, electronic order parameters, and free electron and hole densities~\cite{Shi17Ginzburg,Shi18Phase}. 
It treats the structural distortion and the electron correlation aspects on an equal footing, and has been successfully applied to the determination of the equilibrium stable state under strain/stress and electric field~\cite{Shi17Ginzburg,Shi18Phase}. 
This continuum model is also suitable for describing the kinetics of the IMT in mesoscale systems~\cite{Chen02Phase}. 
In this work, we formulate the dynamical model and apply it to the investigation of the current-driven IMT in VO$_2$. 
To rule out the Joule heating effect and only examine the IMT due to the electron correlation mechanism, we take advantage of the theoretical modeling by considering an isothermal process that may not be readily realized in real experiments. 
We find that the current with only the electron correlation effect can indeed drive the few-nanosecond ultrafast switching. 
The temperature-current phase diagram is further calculated. 
We also find that the current can drive domain walls to move.

\section{Method}
The thermodynamics of the IMT in VO$_2$ can be described by a Landau potential functional (Gibbs free energy) incorporating a contribution from intrinsic VO$_2$ and that from additional free carriers (which may be introduced by doping and electric field)~\cite{Shi17Ginzburg,Shi18Phase},
\begin{align*}
G_\text{t}[T,\Phi;\{\eta_i\},\{\mu_i\},n,p]=& G_0[T;\{\eta_i\},\{\mu_i\}] \notag \\
&+G[T,\Phi;\{\mu_i\},n,p].
\end{align*}
Here $T$ is the temperature, $\Phi$ is the electric potential, $\eta_i,i=1,2,3,4$ are the structural order parameter fields, $\mu_i,i=1,2,3,4$ are the spin-correlation order parameter fields (characterizing the magnetic order), and $n$ and $p$ are the free electron and hole density fields (per unit cell), respectively. 
$\eta_i$ and $\mu_i$ explicitly characterize the structural and the electronic phase transitions during the IMT, respectively: a finite $\eta_i$ indicates the dimerization of the neighboring V atoms, and a finite $\mu_i$ indicates the formation of the dynamical singlet situated on the neighboring V sites and consequently the opening of the energy gap~\cite{Biermann05Dynamical,Zheng15Computation,Brito16Metal}. 
The order parameters of the different phases are: $\eta_1=\eta_3\neq 0, \eta_2=\eta_4=0, \mu_1=\mu_3\neq 0, \mu_2=\mu_4=0$ (and other symmetry-related values) for the M1 phase, $\eta_1\neq 0, \eta_2=\eta_3=\eta_4=0, \mu_1\neq 0, \mu_2=\mu_3=\mu_4=0$ (and other symmetry-related values) for the M2 phase, and $\eta_i=0,\mu_i=0,i=1,2,3,4$ for the R phase~\cite{Shi17Ginzburg}.
The intrinsic Landau potential $G_0$ consists of a bulk energy term and a gradient energy term, and its detailed form can be found in the references~\cite{Shi17Ginzburg,Shi18Phase}. 
In the previous work~\cite{Shi18Phase} we employed the Boltzmann statistics commonly used in semiconductor physics as an approximation to the Fermi statistics for free electrons and holes. 
To better characterize the kinetics of the free electrons and holes, we improve the model by directly using the Fermi distribution to calculate the free electron and hole densities and their Gibbs free energy $G$, although it will bring extra complexity to the modeling.

Since the energy gap opens nearly symmetrically with respect to the Fermi level of the R phase during the metal-to-insulator transition~\cite{Miller12Unusually}, we can set the energy reference to the midpoint of the gap, to simplify the description of the theory. 
With this reference and the simplification of one effective parabolic band for each of the conduction and the valence bands, the electron and hole densities can be written as
\begin{subequations}
\begin{align}
n&=N_\text{c} F_{1/2}\left(\frac{\xi_\text{e}-E_\text{g}/2+e\Phi}{k_\text{B}T}\right), \\
p&=N_\text{v} F_{1/2}\left(\frac{\xi_\text{h}-E_\text{g}/2-e\Phi}{k_\text{B}T}\right).
\end{align}
\label{eq:np}
\end{subequations}
Here the function $F_{1/2}(x)\equiv (2/\sqrt{\pi})\int_0^\infty\sqrt{\epsilon}[1+\exp(\epsilon-x)]^{-1}d\epsilon$ is the Fermi integral. 
$k_\text{B}$ is the Boltzmann constant and $e$ is the elementary charge.
$N_\text{c}=2(m_\text{e}^*k_\text{B}T/2\pi\hbar^2)^{3/2}$ and $N_\text{v}=2(m_\text{h}^*k_\text{B}T/2\pi\hbar^2)^{3/2}$ are the effective densities of states of the conduction band and the valence band, respectively, where $m_\text{e(h)}^*$ is the effective mass of the electrons (holes) and $\hbar$ is the Planck constant over $2\pi$~\cite{Moll64Physics}.
$\xi_\text{e}$ and $\xi_\text{h}$ are the (quasi-) chemical potentials of the electrons and the holes, respectively.
$E_\text{g}$ is the gap and may be directly related to the spin-correlation order parameters~\cite{Biermann05Dynamical,Zheng15Computation,Brito16Metal} $E_\text{g}(\{\mu_i\})\approx 2U^2\mu_0^2\sum_i\mu_i^2/k_\text{B}T_\text{c}$ ($U$ is the onsite Coulomb repulsion and $\mu_0$ is a dimensionless parameter)~\cite{Shi17Ginzburg,Shi18Phase}. 

The Gibbs free energy of the free electrons and holes is then just
\begin{equation*}
G=\int(n\xi_\text{e}+p\xi_\text{h})\frac{dV}{V_0}-G_\text{i}[T;\{\mu_i\}],
\end{equation*}
and using Eq.~(\ref{eq:np}) to eliminate the chemical potentials in it, one obtains
\begin{align}
G=\int\bigg\{ & k_\text{B}T\left[nF_{1/2}^{-1}\left(\frac{n}{N_\text{c}}\right)+pF_{1/2}^{-1}\left(\frac{p}{N_\text{v}}\right)\right]  \notag  \\
& + \frac{E_\text{g}}{2}(n+p)+e\Phi(p-n) \bigg\} \frac{dV}{V_0} - G_\text{i}[T;\{\mu_i\}].
\label{eq:G}
\end{align}
Here $F_{1/2}^{-1}$ represents the inverse function of $F_{1/2}$, and we note that the only term that depends on $\mu_i$ in the integrand is the second one (i.e., the $E_\text{g}$ term).
$G_\text{i}$ is the equilibrium intrinsic Gibbs free energy of the electrons and holes, which makes $G$ vanish, and thus $G_\text{t}$ recover to $G_0$, at equilibrium and zero electric field.
$G_\text{i}$ may have a complicated form.
However, what is directly needed in the simulation is not $G_\text{i}$ itself, but $\delta G_\text{i}/\delta \mu_i$ [see Eq.~(\ref{eq:allen})].
For the latter it can be proven (see Appendix\ref{sec:app} for the derivation) that $\delta G_\text{i}/\delta \mu_i = n_\text{i}d E_\text{g}/d \mu_i$, where $n_\text{i}=N_\text{c}F_{1/2}[(\xi_\text{eq}-E_\text{g}/2)/k_\text{B}T]$ is the intrinsic carrier density ($\xi_\text{eq}$ is the equilibrium intrinsic chemical potential of the electrons).
$dV$ is the infinitesimal volume element and $V_0$ is the unit cell volume. 

The kinetics of the phase transition is described by the Allen-Cahn equations for the non-conserved order parameters $\eta_i$ and $\mu_i$~\cite{Chen02Phase},
\begin{subequations}
\begin{align}
\frac{\partial\eta_i}{\partial t}&=-L_\eta\frac{\delta G_\text{t}}{\delta \eta_i},  \\
\frac{\partial\mu_i}{\partial t}&=-L_\mu\frac{\delta G_\text{t}}{\delta \mu_i},
\end{align}
\label{eq:allen}
\end{subequations}
and the Cahn-Hilliard equations (diffusion equations) for the conserved order parameters $n$ and $p$~\cite{Chen02Phase},
\begin{subequations}
\begin{align}
\frac{\partial n}{\partial t}&=\nabla\cdot\left(\frac{M_\text{e}n}{e}\nabla\frac{\delta G_\text{t}}{\delta n}\right)+s,  \\
\frac{\partial p}{\partial t}&=\nabla\cdot\left(\frac{M_\text{h}p}{e}\nabla\frac{\delta G_\text{t}}{\delta p}\right)+s,
\end{align}
\label{eq:hilliard}
\end{subequations}
where $t$ is the time, $L_\eta$ and $L_\mu$ are constants related to the interface mobilities, $M_\text{e(h)}$ is the electron (hole) mobility, and $s$ is the source term representing the electron-hole recombination process. 
Note for the thermodynamic relations $\delta G_\text{t}/\delta n=\xi_\text{e}$ and $\delta G_\text{t}/\delta p=\xi_\text{h}$.

The source term may have the form $s=K(\{\mu_i\})(n_\text{eq}p_\text{eq}-np)$, where $n_\text{eq}=N_\text{c}F_{1/2}[(\xi_\text{eq}-E_\text{g}/2+e\Phi)/k_\text{B}T]$ and $p_\text{eq}=N_\text{v}F_{1/2}[(-\xi_\text{eq}-E_\text{g}/2-e\Phi)/k_\text{B}T]$ are the equilibrium densities of the electrons and the holes, respectively, and $K$ is the recombination rate coefficient independent of $n$ and $p$. 
In the insulating phase, $K$ is finite.
In the metallic phase, however, $K$ should be zero: the holes appearing in the metallic phase in the model are not the genuine holes as in the insulating phase, but rather should be interpreted as an effective positive-charge background for the free electrons as to achieving charge neutrality, in which case the concept of the electron-hole recombination is not applicable.
To account for this, we assume the symmetry-allowed lowest order dependence of $K$ on the electronic order parameters, $K=K_0\sum_i\mu_i^2$, where $K_0$ is a constant.  

Equations~(\ref{eq:allen},\ref{eq:hilliard}) are closed by the Poisson equation for the self-consistent determination of the electric potential $\Phi$,
\begin{equation*}
-\nabla^2\Phi=\frac{e(p-n)}{\epsilon_0\epsilon_\text{r}},
\end{equation*}
where $\epsilon_0$ and $\epsilon_\text{r}$ are the vacuum dielectric permittivity and the relative dielectric permittivity of VO$_2$, respectively. 
In the simulations, for Eq.~(\ref{eq:hilliard}) we use the energies $\gamma_\text{e}\equiv \xi_\text{e}-E_\text{g}/2+e\Phi$ and $\gamma_\text{h}\equiv \xi_\text{h}-E_\text{g}/2-e\Phi$ as the unknown variables instead of $n$ and $p$, and obtain $n$ and $p$ through Eq.~(\ref{eq:np}) after solving for $\gamma_\text{e}$ and $\gamma_\text{h}$.

\begin{figure}[t!]
 \includegraphics[width=0.37\textwidth]{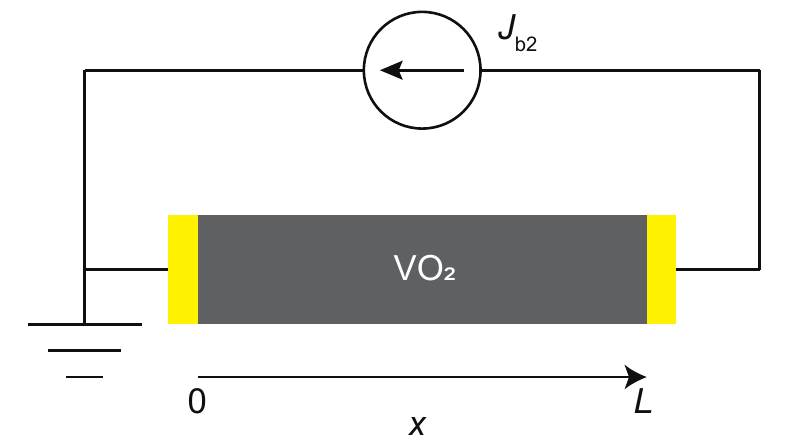}
 \caption{\label{fig:geo} Schematics of the geometry used in the simulations. $L$ is the length of the VO$_2$ sample, and is set to $100$~nm in the simulations. The gold region represents the electrode.}
\end{figure}

The boundary conditions are schematically represented in Fig.~\ref{fig:geo}. 
The left boundary ($x = 0$) is connected to the ground, i.e., we have
\begin{align*}
\Phi|_{x=0}&=0, \\
\gamma_\text{e}|_{x=0}&=\gamma_\text{h}|_{x=0}=\gamma_\text{b1},
\end{align*}
where $\gamma_\text{b1}$ is a constant corresponding to a fixed carrier density $n_\text{b1}$ at the boundary, $n|_{x=0}=p|_{x=0}=n_\text{b1}$.
The right boundary ($x = L$) has a constant flux. 
We assume the boundary condition for $\Phi$ at $x = L$ to correspond to a small constant electric field in the electrode $E_\text{lctrd2}$ (we set $E_\text{lctrd2}$ to $0.001$~MV/m). 
Eventually we have,
\begin{align*}
(\partial_x\Phi)|_{x=L}+E_\text{lctrd2}&=\frac{e(p-n)|_{x=L}\lambda}{\epsilon_0\epsilon_\text{r}}, \\
j_\text{e}|_{x=L}=-\frac{J_\text{b2}}{e}&,~j_\text{h}|_{x=L}=0,
\end{align*}
where $\lambda$ is the length of the charge depletion region at the boundary and is set to $5$~nm, $j_\text{e}=-(M_\text{e}n/e)\partial_x\xi_\text{e}$ [$j_\text{h}=-(M_\text{h}p/e)\partial_x\xi_\text{h}$] is the electron (hole) flux, and $J_\text{b2}$ is the constant boundary current density. 
In the simulations we find that different values of $E_\text{lctrd2}$ and $\lambda$ have minor influence on the results.
We assume zero flux for the order parameters $\eta_i$ and $\mu_i$ at both boundaries, i.e., $(\partial_x\eta_i)|_{x=0,L}=(\partial_x\mu_i)|_{x=0,L}=0$, which corresponds to no interaction of the order parameters at boundaries.

\begin{table}[b!]
\caption{\label{tab:para}Values of the parameters adopted from experiments. $m_\text{e}$ is the electron mass.}
\begin{ruledtabular}
\begin{tabular}{ccccccc}
\makecell{$m_\text{e,h}^*$ \\ ($m_\text{e}$) \\ \cite{Fu13Comprehensive}} & \makecell{$M_\text{e}$ \\ (cm$^2$/Vs) \\ \cite{Rosevear73Hall}} & \makecell{$\dfrac{M_\text{e}}{M_\text{h}}$ \\ \cite{Miller12Unusually}} & \makecell{$\tau_\text{eh}$ \\ ($\mu$s) \\ \cite{Miller12Unusually}} & \makecell{$\tau_\eta$ \\ (ps) \\ \cite{Cavalleri01Femtosecond}} & \makecell{$\tau_\mu$ \\ (fs) \\ \cite{Wegkamp14Instantaneous}} & \makecell{$\epsilon_\text{r}$ \\ \cite{Yang10Dielectric}}  \\
\hline
65 & 0.5 & 1.2 & 10 & 1 & 10 & 60 \\
\end{tabular}
\end{ruledtabular}
\end{table}

We estimate the parameters in the model based on experimental results. 
To our best knowledge, the hole mobility in VO$_2$ has not yet been directly measured. Nonetheless, we estimate the ratio of the electron and hole mobilities $M_\text{e}/M_\text{h}\approx 1.2$ from the position of the photocurrent peak in the scanning photocurrent microscopy measurement~\cite{Miller12Unusually}. 
The constant characterizing the electron-hole recombination rate $K_0$ can be calculated from the free carrier lifetime $\tau_\text{eh}\sim 10$~$\mu$s~\cite{Miller12Unusually} through the relation $K_0 = (2n_\text{ic}\tau_\text{eh})^{-1}$~\cite{Shockley52Statistics}, where $n_\text{ic}$ is the intrinsic carrier density of the insulating phase near $T_\text{c}$ (note that $\sum_i\mu_i^2 \sim 1$ in the insulating phase). 
Similarly, $L_\eta$ and $L_\mu$ can be estimated from the characterization times of the structural and the electronic phase transitions $\tau_\eta \sim 1$~ps~\cite{Cavalleri01Femtosecond} and $\tau_\mu \sim 10$~fs~\cite{Wegkamp14Instantaneous}, by $L_\eta \sim [\tau_\eta a(T_\text{c}-T_\eta)/T_\text{c}]^{-1}$ and $L_\mu \sim (4U^2\mu_0^2n_\text{ex}\tau_\mu/k_\text{B}T_\text{c})^{-1}$, respectively. 
Here $a$ and $T_\eta$ are the Landau coefficient and the Curie-Weiss temperature of the quadratic term of $\eta_i$, respectively (see~\cite{Shi17Ginzburg,Shi18Phase}), and $n_\text{ex} \approx 0.08$ per V atom is the photoexcited free electron density in the measurement of $\tau_\mu$~\cite{Wegkamp14Instantaneous}. 
The values of the parameters adopted from experiments are summarized in Table~\ref{tab:para}.

\section{Current-driven ultrafast switching and phase diagram}

\begin{figure}[t!]
 \includegraphics[width=0.48\textwidth]{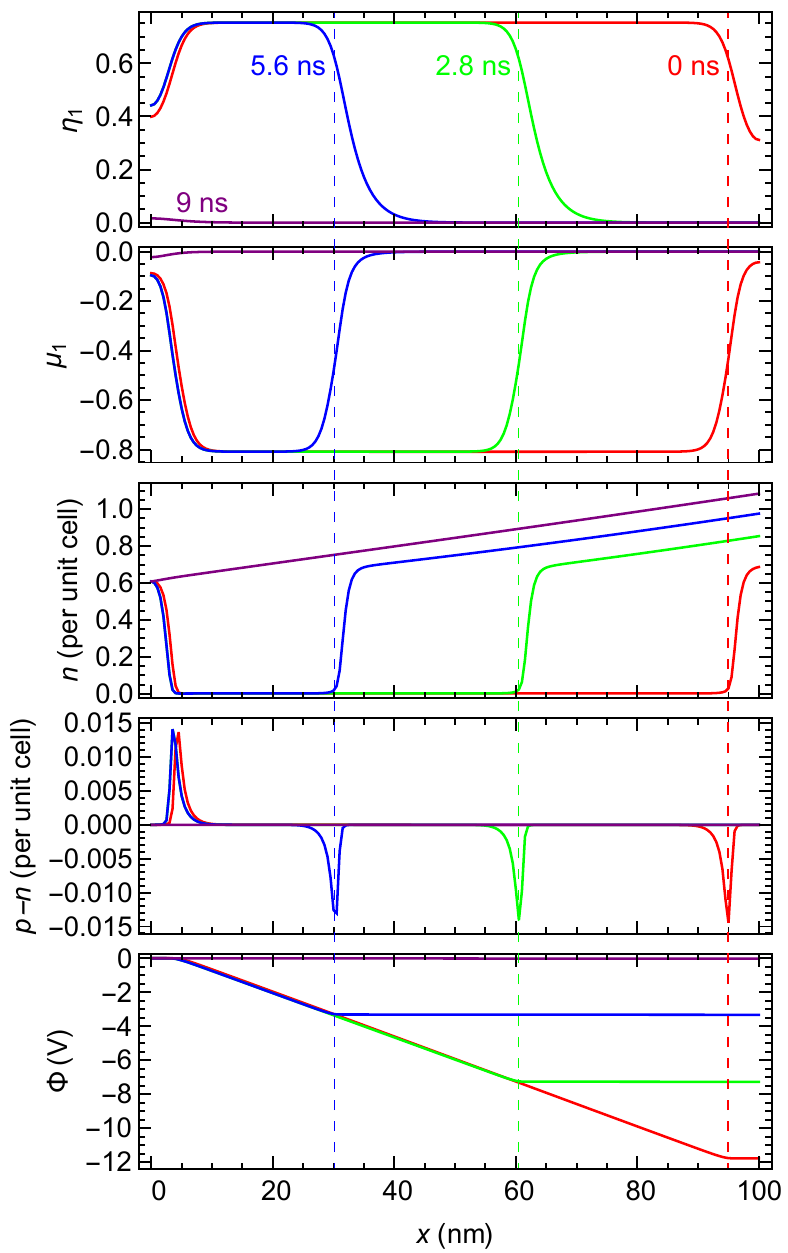}
 \caption{\label{fig:ultfas} Simulated temporal evolution of various variables during the current-driven ultrafast switching in VO$_2$ at $T=320$~K, $J_\text{b2}=57.8$~nA/nm$^2$ and $n_\text{b1}\approx 0.6$ per unit cell. During the process, $\eta_3$ ($\mu_3$) is the same as $\eta_1$ ($\mu_1$), and $\eta_2=\eta_4=0$ and $\mu_2=\mu_4=0$. The dashed lines indicate the positions of the insulator-metal interface at different times.}
\end{figure}

We first investigate the case in which the VO$_2$ sample has an initial equilibrium M1 phase in the bulk and is subject to a large current density $\sim 10^1$ nA/nm$_2$. 
This could be the case in the measurements of the voltage-pulse-induced ultrafast switching in VO$_2$~\cite{Leroy12High,Zhou13Voltage,Chae05Abrupt}. 
Figure~\ref{fig:ultfas} shows the calculated temporal evolution of various variables at $T = 320$~K, $J_\text{b2} = 57.8$~nA/nm$^2$ and $n_\text{b1} \approx 0.6$ per unit cell. 
We find that $n_\text{b1}$ has minor influence on the profiles of the variables in the bulk and the switching time. 
At $t = 0$~ns, the structural order parameters and the electronic order parameters have uniform equilibrium finite values $\eta_1 = \eta_3 = 0.76$ and $\mu_1 = \mu_3 = -0.84$ in the bulk ($\eta_2 = \eta_4 = 0$ and $\mu_2 = \mu_4 = 0$), indicating the initial state is a uniform monoclinic insulator (M1 phase). 
$\eta_1$ ($\eta_3$) and $\mu_1$ ($\mu_3$) then turn to zero from the $x = L$ end, representing that the rutile metal (R phase) grows from the $x = L$ end. 
This is in contrast to the Joule-heating-induced switching, in which the initial insulator turns into the metal uniformly due to the uniform heating. 
The metallic phase spreads from the $x=L$ end to the $x=0$ end in $\sim 9$~ns, yielding a few-nanosecond ultrafast switching. 
This reconciles with the few to tens of nanoseconds switching time found in the voltage-pulse-induced IMT in VO$_2$~\cite{Leroy12High,Zhou13Voltage,Chae05Abrupt}.

The growth of the metallic phase from the $x = L$ end is driven by the carrier doping from the carrier injection and the negative electric potential~\cite{Nakano12Collective,Shi18Phase} at that end. 
The excess carriers screen the electron-electron repulsion and thus reduce the electron correlation, thereby stabilize the metallic phase~\cite{Kim04Mechanism,Stefanovich00Electrical,Wegkamp14Instantaneous}. 
As the metallic phase grows, there are net negative charges accumulating at the insulator-metal interface and following it. 
The electric potential becomes flat inside the metallic phase, as it should.

\begin{figure}[t!]
 \includegraphics[width=0.4\textwidth]{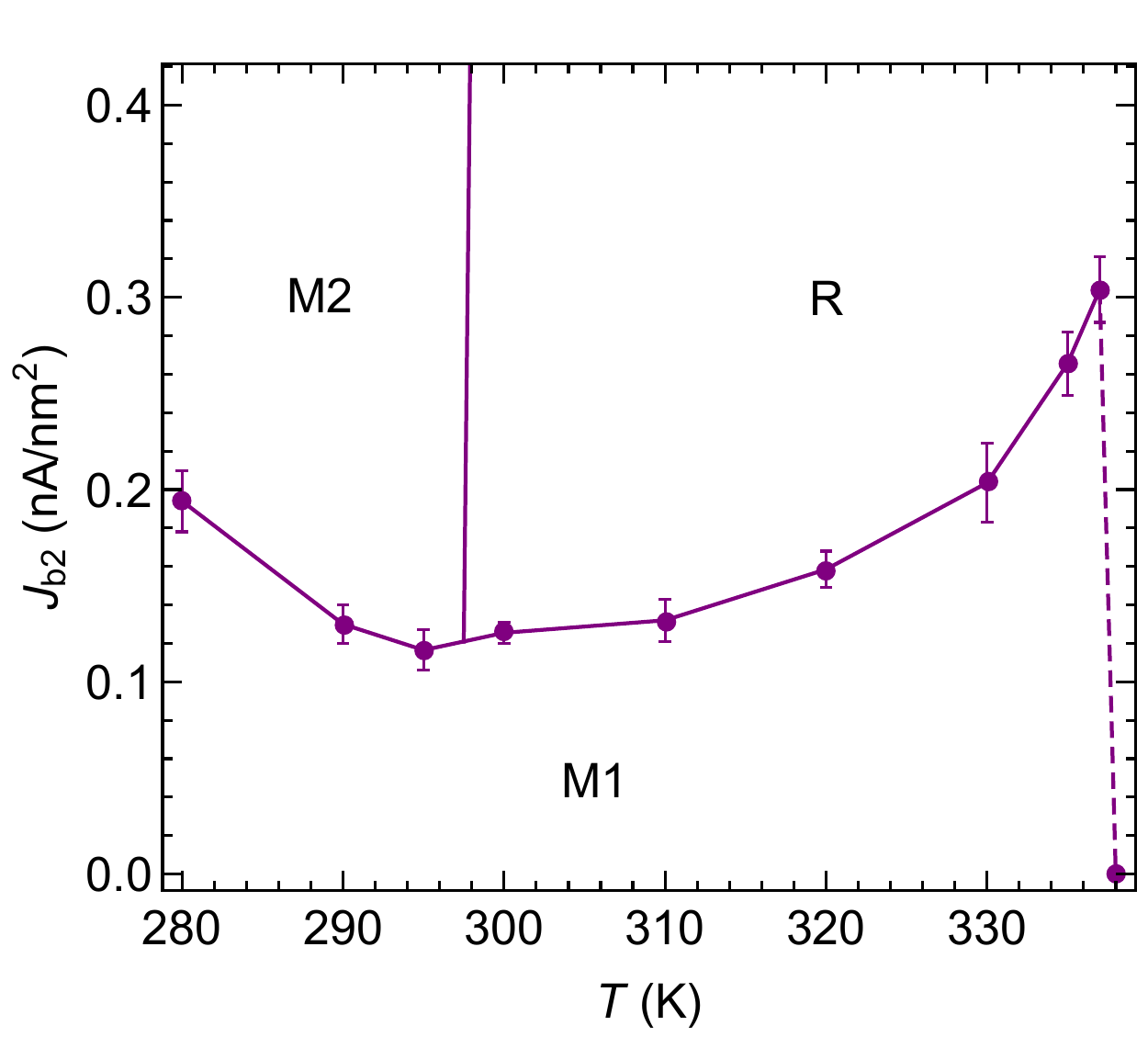}
 \caption{\label{fig:pd} Calculated temperature versus current density phase diagram of VO$_2$. The dots with error bars are the calculated points on the phase boundaries, and the lines are guide to eyes. The M2-R phase boundary has a large slope: the critical current density at $300$~K of the M2-R transition is calculated to be $2.1\pm 0.2$~nA/nm$^2$ (which is not included in the plot range for clarity). The dashed line represents the discontinuous point.}
\end{figure}

Knowing that the current can induce the IMT isothermally, we further calculate the temperature versus current density phase diagram of VO$_2$ under isothermal condition. 
The result is shown in Fig.~\ref{fig:pd}. 
It should be noted that any point on the phase diagram corresponds to a nonequilibrium steady state, not an equilibrium stable state. 
Strikingly, the simulation suggests that the current may induce the M2 phase at low temperatures ($< 298$~K). 
The M2-R phase boundary has a large positive slope, indicating that a current with large enough density may eventually drive the M1 phase to the M2 phase even at high temperatures ($> 298$~K).

The calculated critical current density of the M1-R phase transition is comparable with the experimentally measured values with the Joule heating effect present ($10^{-2} \sim 10^{-1}$~nA/nm$^2$)~\cite{Zimmers13Role,Joushaghani14Voltage}. 
This confirms that the Joule heating effect and the electron correlation effect are indeed deeply entangled in the current-driven IMT. 
On the other hand, the critical current density of the M1-R phase transition increases at elevating temperature, contrary to the Joule-heating-induced IMT in which the critical current density naturally decreases at elevating temperature. 
This finally leads to the presence of a discontinuous point at $T_\text{c}$, as shown by the dashed line in Fig.~\ref{fig:pd}. 
The discontinuity in phase diagrams is abnormal, however it may be the case in this anomalous phase diagram that corresponds to nonequilibrium steady states.

\section{Current-driven domain wall motion}

\begin{figure}[t!]
 \includegraphics[width=0.48\textwidth]{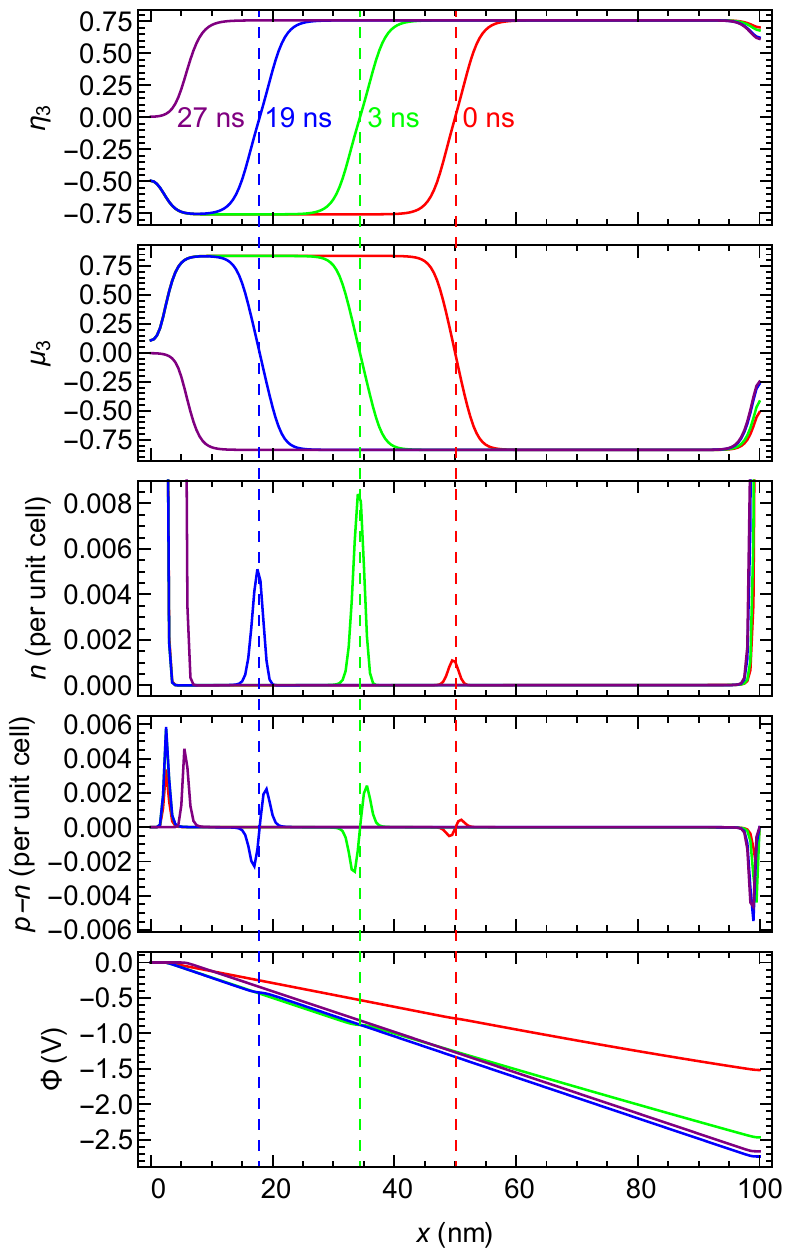}
 \caption{\label{fig:dwmot} Simulated temporal evolution of various variables during the  current-driven domain wall motion in VO$_2$ at $T=320$~K, $J_\text{b2}=0.0811$~nA/nm$^2$ and $n_\text{b1}\approx 0.6$ per unit cell. During the process, $\eta_1$ ($\mu_1$) has a nearly uniform value $0.76$ ($-0.84$) along the sample despite at the boundaries, and $\eta_2=\eta_4=0$ and $\mu_2=\mu_4=0$. The dashed lines indicate the positions of the twin wall within the M1 phase at different times. The range of the finite net charge region at $x=L$ boundary is within $\lambda\triangleq 5$~nm, which justifies this setting of $\lambda$.}
\end{figure}

We then examine how the current affects the domain wall in VO$_2$. 
The initial configuration is set to a two-domain structure within the M1 phase, with the domain wall (twin wall) located at $x = L/2$. 
This is shown by the profiles of $\eta_3$ and $\mu_3$ at $t = 0$~ns in Fig.~\ref{fig:dwmot}. 
The order parameters of the right domain are $\eta_1 = \eta_3 = 0.76, \eta_2 = \eta_4 = 0, \mu_1 = \mu_3 = -0.84, \mu_2 = \mu_4 = 0$, which is denoted as the variant 1 of the M1 phase. 
The order parameters of the left domain are $\eta_1 = -\eta_3 = 0.76, \eta_2 = \eta_4 = 0, \mu_1 = -\mu_3 = -0.84, \mu_2 = \mu_4 = 0$, which corresponds to a $180^\circ$ rotation about the rutile $c$ axis of the variant 1, and is denoted as the variant 3 of the M1 phase. 
As can be seen in Fig.~\ref{fig:dwmot}, upon the application of a current with a small density (not adequate to trigger the IMT), the twin wall between the variant 1 and the variant 3 moves opposite to the current direction (i.e., $-x$ direction), and finally moves to the $x = 0$ end in $27$~ns, leading to the vanishing of the variant 3.

The twin wall has a relatively large carrier density, and thus a relatively large conductivity compared to the interior of the domains. 
The net charges localized at the twin wall form an effective dipole oriented along the direction of the electric field. 

We note that this current-driven twin wall motion cannot be realized via the Joule heating effect, since the Joule heating effect is symmetric about $\pm x$ directions. 

\section{Conclusion}
We formulated a phase-field model that takes into account the structural distortion, the electron correlation and the free carrier aspects to describe the mesoscale kinetics of the IMT in VO$_2$. 
We applied it to the investigation of the isothermal current-driven IMT in VO$_2$. 
The simulation showed that the current can drive a few-nanosecond ultrafast switching isothermally through the electron correlation effect. 
The temperature versus current density phase diagram was further obtained, which indicates that the current may induce the M2 phase at low temperatures under isothermal condition. 
The current was also shown to be able to drive the domain wall to move, which could potentially be useful such as to conveniently transform a multi-domain sample to a single-domain sample.
Our work may assist related experiments and provide guidance to the engineering of VO$_2$-based electric switching devices.

\begin{acknowledgments}
This work was funded by the Penn State MRSEC, Center for Nanoscale Science, under the award NSF DMR-1420620.
\end{acknowledgments}

\appendix*
\section{derivation of $\delta G_\mathrm{i}/\delta \mu_i$} \label{sec:app}
Let us first denote the integral in Eq.~(\ref{eq:G}) at $\Phi=0$ as $G^0[\{\mu_i\},n,p]$.
Then by definition $G_\text{i}[\{\mu_i\}]=G^0|_{n,p=n_\text{i}}$. 
Next we denote the first term in the integrand in Eq.~(\ref{eq:G}) as $g(n,p)$.
After this preparation, we shall start the derivation.
Since $G_\text{i}$ depends on $\mu_i$ only through $E_\text{g}(\{\mu_i\})$, we obtain
\begin{equation}
\frac{\delta G_\text{i}}{\delta \mu_i} = \frac{\delta G_\text{i}}{\delta E_\text{g}}\frac{dE_\text{g}}{d\mu_i}.
\label{eq:partial1}
\end{equation}
We also have
\begin{align}
\frac{\delta G_\text{i}}{\delta E_\text{g}} &= \frac{\delta G^0}{\delta E_\text{g}}\bigg|_{n,p=n_\text{i}}+\left( \frac{\delta G^0}{\delta n}+\frac{\delta G^0}{\delta p} \right)\bigg|_{n,p=n_\text{i}} \frac{dn_\text{i}}{dE_\text{g}} \notag  \\
&= n_\text{i} + E_\text{g}\frac{dn_\text{i}}{dE_\text{g}} + \left( \frac{\partial g}{\partial n} + \frac{\partial g}{\partial p} \right)\bigg|_{n,p=n_\text{i}}\frac{dn_\text{i}}{dE_\text{g}}.
\label{eq:partial2}
\end{align}
But from the equilibrium condition $(\delta G^0/\delta n)|_{n,p=n_\text{i}} = \xi_\text{e} = \xi_\text{eq}$ and $(\delta G^0/\delta p)|_{n,p=n_\text{i}} = \xi_\text{h} = -\xi_\text{eq}$~\cite{Shi18Phase}, we have
\begin{align*}
\frac{\partial g}{\partial n}\bigg|_{n,p=n_\text{i}} &= \xi_\text{eq} - \frac{E_\text{g}}{2}, \notag  \\
\frac{\partial g}{\partial p}\bigg|_{n,p=n_\text{i}} &= -\xi_\text{eq} - \frac{E_\text{g}}{2}.
\end{align*}
Substituting these two equations into Eq.~(\ref{eq:partial2}), one finds that many terms cancel out, and obtains beautifully
\begin{equation*}
\frac{\delta G_\text{i}}{\delta E_\text{g}} = n_\text{i}.
\end{equation*}
The substitution of this equation in Eq.~(\ref{eq:partial1}) just gives the desired relation
\begin{equation*}
\frac{\delta G_\text{i}}{\delta \mu_i} = n_\text{i} \frac{dE_\text{g}}{d\mu_i}.
\end{equation*}
This completes the proof.

\bibliography{VO2curr_ref}

\end{document}